\documentclass[12pt]{article}
\usepackage{amsmath,euscript,array,amssymb,cite}

\newcommand{\beq}{\begin{eqnarray}}
\newcommand{\eeq}{\end{eqnarray}}
\begin{document}

\baselineskip=18pt

\begin{titlepage}
\begin{center}
         \hfill {}\\
       \hfill {}\\
\hfill{}
\vskip 1.2in

{\LARGE \bf An Invitation to String Theory}

\vskip .5in
{\bf
      Chethan Krishnan}


{\em Department of Physics,
University of Texas at Austin, Austin, Texas, 78712, USA}

{\tt
chethan@physics.utexas.edu}
\end{center}

\vskip .5in
\begin{abstract}
These lectures are supposed to give a point of entry into that vast web of 
related ideas that go under the name ``string theory". I start 
with a more or less qualitative introduction to gravity as a field 
theory and sketch how one 
might try to quantize it. Quantizing gravity using the usual techniques of 
field theory will turn out to be unsuccessful, and that will be a 
motivation for 
trying an indirect approach: string theory. I present 
bosonic string theory, quantize the closed string in the light cone gauge 
and show that one of the states in the closed string Hilbert space can 
be interpreted as a graviton. I will end with some general 
qualitative ideas on slightly more advanced topics.    

\end{abstract}
\end{titlepage}

\newpage

\section{Gravity and the Necessity for String Theory}

I want to thank Prof. Sathish for giving me a chance to speak here about
string theory. I hope I can rise to the expectations that might have been
generated by his extremely kind and generous introduction. The aim of the
organizers to expose the students to topics beyond curricula is truly laudable. Often, the inspiration for trudging through a
lot of what goes for as ``syllabus" is to be found in what lies beyond it.
By giving the impression that whatever that is worth doing has already
been done, and by making it seem that the people who make contributions in
physics are larger than life (and often belong to a different place, time
and race), the educational system fails to impart the most important
ingredient that a theoretical physicist needs: the confidence to write
down an equation. The remedy for this, I believe, is to not present
physics as a completed business, and to expose the inadequacies and
failings of our present world-view. That way, it becomes clear that there
are many profound questions which have remained unanswered despite the 
work of many great people.  

String theory is necessary because we don't know of any other way in which
to make a theory of gravity that doesn't contradict Quantum Mechanics. To
me, this is the most compelling, indeed irresistible, reason for liking
string theory. There is not a single experiment or observation that
violates the principles of QM, and so the construction of a Quantum
theory of Gravity is a necessity, not an option: and this is precisely
what string theory claims to provide. There are some other reasons for
liking string theory - all of them impressive: string theory manages, at
least at a schematic level, to unify\footnote{There seems to be some 
confusion among the students about unification: Grand Unified Theories 
(GUTs) 
are theories that unify all the forces EXCEPT gravity - they are not so 
grand in that sense. But string theory 
claims to go beyond that and includes gravity as well. GUTs are specific 
quantum field theories, just like QED or scalar field theory, but string 
theory is much more than a QFT.} all the forces (gravity being 
just one of the four) in nature. Also we believe that string theory is an
absolutely unique theory - we cannot tweak the parameters in the theory to
make it do what we want; changing things capriciously will make it
mathematically inconsistent. But I feel that these reasons are secondary 
and
more aesthetic. If the forces in nature are unified, thats nice and
pretty, but we have no fully compelling {\it a priori}
 reason for 
believing that
the forces {\it have to} be unified. And again,
uniqueness is wonderful and we would expect that from a final theory, but
how do we know that there is a final theory?  For my part, I do think that
there is one, but I do not know of a watertight argument that would
convince a skeptic.

The theory of gravity that we know and love is Einstein's GR. It has
certain problems of its own, like the existence of singularities - regions
of spacetime where the theory breaks down. A theory that fails at certain
limits of its range of applicability is not what we would call a complete
theory. Usually singularities arise because of the infinite resolution
that a manifold model of spacetime provides. Quantization and the
fuzziness associated with it might be enough to save this problem 
of classical gravity (by which we mean GR).

The historical origins of GR are very different from that of the field
theories of other forces. I want to look at gravity from a purely field
theoretic perspective because the only way in which we know  how to make
consistent quantum theories of interactions is through quantizing fields.
So I will pretend that we don't know about GR, and we want to make a field
theory of gravity based on the known features of gravitation and the
general principles of field theory\footnote{This path was first tried by 
Gupta 
and later by Feynman, and brought to completion by Deser
\cite{Deser:1969wk, Boulware:1974sr, 
Deser:1987uk}.}. Remarkably, 
we will be led to GR.
Then I will try to show that the general techniques of field theory imply
that there are serious problems in quantizing this theory. By failing to
quantize gravity face-on, I hope to make indirect approaches seem more
plausible. String theory is such an indirect approach to incorporate
gravity in a quantum framework.

In the first few sections, we will only need general arguments to drive us 
to our conclusions. It is not very meaningful to try to be too exacting 
in obtaining a theory of gravity which we know is imperfect. So 
the reader should focus only on the general line of reasoning and not 
focus too much on the nitty-gritty. Even though the mathematical level 
required in the first few sections is minimal, a certain maturity with 
field theory might be necessary. Later, the parts on string theory are 
pretty much independent of these initial sections, so one should not be 
too bothered by an occasional unclear point. 

\section{Gravity as a Field Theory}

The first general principle that I will use in constructing a field theory
of gravity is that my theory has to respect special relativity - it should
be Lorenz covariant. Mathematically, ``covariant" means that the Equations
of Motion (EOM) in the theory (the ``Laws of Nature") should transform
under some representation of the Lorentz group, SO(3,1). This can be
accomplished by taking the fields themselves to form representations, and
constructing the action functional out of them as a scalar. This way, the
EOM that results from varying the action with respect to the fields is
automatically covariant. So, if I want to make my theory respect special
relativity, I will choose my fields in some Lorentz representation. It is
a well-known mathematical fact that a number, called the spin of the 
field,
can label representations of the group SO(3,1). This number can only be
integral (0, 1, 2,...) or half-integral (1/2, 3/2,...). So this gives us a 
list of possible fields in any Lorentz covariant field 
theory\footnote{The integer spin Lorentz representations are often called 
tensors - in particular spin 0 is called a scalar, and 
spin 1 is called a vector.}. The question 
we should answer is which choice of spin corresponds to the graviton.

Before proceeding further lets ask whether the graviton is a massive field. 
We know from Newtonian gravity that the potential between two masses falls 
off as $1/r$. Now, in QFT, one can calculate the potential due to an 
interaction by comparing the tree-level amplitude in field 
theory with the Born approximation - I will skip the details, since only 
the result is of interest to us. The result is 
that the potential due to the exchange of a field of mass $m$ is 
$\sim e^{-mr}/r$, a Yukawa potential. If this has to reproduce the 
Newtonian 
result we see that the graviton has to be massless.

One of the first things we notice about gravity is that 
macroscopic gravitational 
field configurations can have classical descriptions: we can measure the 
field strengths at different points using classical test masses. This is 
analogous 
to electromagnetism where the electric field strength can be measured 
by introducing test charges. Now, a classical field 
configuration consists of a superposition of states with a large number 
of field quanta\footnote{If the occupancy of the states is not 
large, then the fluctuations in the field strength will be significant 
and we will not have a classical field.}. That is, a 
classical gravitational field is made of states with an enormous number 
of gravitons, just like a classical electromagnetic field is made of 
states with an enormous number of photons. Now comes Pauli's 
celebrated Spin-Statistics theorem. It implies 
that particles that arise as the quanta of fields of half-integral spins 
have the property that {\it not} more than one particle can be put into 
any 
state. So the half-integral spin fields cannot form classical field 
configurations\footnote{The 
inability of electrons (spin 1/2) to form classical field distributions is 
the reason why we think of them classically as particles, not fields. Vice versa for 
the spin 1 electromagnetic field.}. A 
direct consequence of this fact is that we can immediately rule out all 
half-integral spin fields as candidates for a graviton because we know 
that classical 
gravitational fields exist.   

So now our attention is limited to the integral spin representations of 
the Lorentz group. It is known that it is very 
hard to construct consistent interacting quantum field theories with 
particles of high spin, so we will start with the lowest spins 
and hope that we will find a suitable graviton from among one of the low 
spin representations. 

{\bf Scalar $\phi$ (spin 0):} The graviton cannot be a scalar field 
because it is 
impossible to construct a gauge invariant coupling of the scalar with the 
electromagnetic field: the simplest possibility is $\phi A^{\mu}A_{\mu}$, 
which is obviously not gauge invariant because of the explicit appearance 
of the vector potential.  Gauge invariance of electromagnetism is 
necessary for its consistency, so we cannot sacrifice that. But without a 
coupling between gravity and the electromagnetic field, scalar theory of 
gravity will predict no deflection of light by a massive star; contrary to 
the solar eclipse observations.
 
Another possibility is to try something like $\phi F^{\mu \nu}F_{\mu 
\nu}$ which 
is gauge invariant. But here the problem is that this term has too many 
derivatives ($F_{\mu \nu}=\partial_{\mu}A_{\nu}-\partial_{\nu}A_{\mu}$), 
and so the coupling constant for this term in the action must have 
negative mass dimension, hence it is an unrenormalizable interaction which 
is irrelevant (in the Renormalization Group sense) at low energies. But light 
deflection by massive objects is a low energy phenomenon. 

So we have to give up the scalar as a candidate graviton.

{\bf Vector $B_{\mu}$ (spin 1):} A typical example of a massless vector 
field theory is QED. The problem here is that the theory as we all know, 
gives rise to two kinds
of interactions - attractive or repulsive depending upon the charges.
But gravity is universally attractive and so vector field cannot work.

{\bf 2-tensor $h_{\mu\nu}$ (spin 2):} This is the one remaining 
possibility with 
any hope of salvation! If we go any higher in spin, the theories are 
pathological. There are no 
obvious problems with $h_{\mu\nu}$ as the graviton. So lets see how far we 
can proceed. First, as always in perturbative field theory, we will write 
down the free field theory and then 
we 
will add interactions. The action for a free $h_{\mu\nu}$ theory I 
will write 
schematically as, 
\beq
\label{free gravity}
S\sim \frac{1}{2}\int d^4 x (\partial_{\mu} 
h^{\rho\nu}\partial^{\mu}h_{\rho\nu}). 
\eeq
There can be other terms with other possible contractions between indices, 
but we are interested only in the general idea, so we won't bother 
about the details inconsequential to us. Notice that in the above 
action, there is no mass term since the graviton is massless. 

We know that gravity couples to anything 
that has energy, so when we couple $h_{\mu\nu}$ to matter the 
natural coupling to try is $h_{\mu\nu}T^{\mu\nu}$, where $T^{\mu\nu}$ is 
the energy-momentum tensor which can be taken to be symmetric. This 
implies that $h_{\mu\nu}$ is symmetric as well. 

Gravity couples to energy, but the gravitational field 
itself has energy: using the usual canonical procedures, we can define a 
(positive semidefinite) Hamiltonian 
starting with the action 
for the gravitational field. If this Hamiltonian is not zero\footnote{Actually what I should really be saying is that the 
Hamiltonian should be zero on the constraint surface in phase space, in 
the sense of Dirac's constraint theory; but this is a technical detail 
that I do not wish to get into.}, then 
it means that not all the energy in the gravitational field couples back 
onto the 
field. The way to make sure that there is no energy 
left uncoupled to the gravitational field is to add nonlinear field 
self-interaction terms in the free field action so that the Hamiltonian 
calculated from the final action is identically zero.

Now, Weinberg has proved a theorem\footnote{The proof will take us too 
far afield.} which says 
that if we start 
with $h_{\mu\nu}$ and impose the condition that it couple consistently 
to $T^{\mu\nu}$, it should have a gauge invariance $h_{\mu\nu}\rightarrow 
h_{\mu\nu}+\partial_\mu \xi_\nu +\partial_\nu \xi_\mu$. Also, if we try to 
add nonlinearities in the free theory action so that we try to preserve 
this gauge invariance, we are led uniquely\footnote{upto terms with more 
derivatives that are suppressed at low energies.} to the Einstein-Hilbert 
action:
\beq
\label {E-H}
S=\int d^4 x \sqrt{-g} R + \kappa S_{matter},
\eeq  
where $g_{\mu\nu}\equiv\eta_{\mu\nu}+h_{\mu\nu}$ and the Ricci scalar is 
calculated from $g_{\mu\nu}$. The coupling constant $\kappa$ is related to 
Newton's constant by $\kappa = 8\pi G/ c^4$. 

So we see that even starting with the principles of relativistic field 
theory, we are driven to Einstein's theory of gravity. So the question of 
quantizing gravity is unavoidably the question of finding a quantum 
generalization of GR. 

\section{Problems in Quantizing GR}

First thing one must notice is that though the expression {\it 
looks} compact, 
the Einstein-Hilbert action is actually extremely complex and highly 
nonlinear if written in terms 
of the basic field variable, namely the metric. Remember that 
\beq
R=g^{ik}(\Gamma^l_{il,k}-\Gamma^l_{ik,l}+\Gamma^m_{in}\Gamma^n_{mk}-\Gamma^m_{ik}\Gamma^n_{mn}) 
\eeq
where
\beq
\Gamma^i_{jk}=\frac{1}{2}g^{im}(g_{mj,k}+g_{mk,j}-g_{jk,m}).
\eeq
The standard procedure for dealing with nonlinear field theories is 
perturbation theory. The philosophy is that we look at the ``free" 
part of the action, by 
which we mean only terms quadratic in the fields, 
and quantize this free field action using the usual techniques of free 
field theory. Then we 
calculate the amplitude for processes in the full nonlinear theory as a 
perturbation series. The basic idea of perturbation theory is that there 
are many ways for a process to take place when there are interactions 
(i.e, nonlinearities), 
and we have to sum over all these possibilities to get the total amplitude 
for the process. Basically this results in sums over intermediate states, 
and if we are 
working in momentum space, these sums become integrals over all 
momenta, $p$. 
It so happens that sometimes these integrals diverge in the $p \rightarrow 
\infty$ limit and we say that the theory has ultraviolet divergences. 
Perturbation theory works under the premise that the higher order terms 
are smaller than the lower order terms, and if a higher order term 
diverges, we are in trouble. Still, in certain restricted classes of 
theories we can make sense out of perturbation theory by rendering the 
infinities harmless by a process called renormalization. The quantum field 
theories that we use for explaining the forces in the standard model are 
all renormalizable. So when we try to quantize gravity using perturbation 
theory, we must address the question whether gravity is renormalizable, 
otherwise the approach doesn't make sense. 

The theory of renormalization is rather elaborate, but there is a simple 
check to see whether a theory is renormalizable or not. Before 
elaborating on what that test is, I should make a small digression on 
dimensional analysis. In field theory, we choose to work in units such 
that 
$\hbar=c=1$. $c=1$, implies that $[L]\sim [T]$ and  
$\hbar=1$ means $[M][L]^2[T]^{-1}\sim 1$. These two together lead to 
$[M]\sim [L]^{-1}$. One must understand that setting $\hbar=c=1$ is not 
some new profound insight, but just means that we just are willing to give 
up with some information that we have at our disposal. We just {\it 
choose} not to distinguish between certain dimensions (for example, length 
and time are seen as the same thing and so is mass and inverse 
length). This is basically a statement that we do not keep track of the 
difference, not that there is no difference. 

Now we can state the condition for a theory to be renormalizable: a theory
is renormalizable if the power of the mass dimension of the coupling
constant in the theory is non-negative. If its negative, its
non-renormalizable, and we have a problem. So is gravity renormalizable?  
Well, to check that we have to check the dimension of the Gravitational
coupling constant, the constant that comes up in the Einstein-Hilbert
action. Since $c=1$, this is nothing but Newton's constant, $G$. From 
Newton's law of
gravitation we can immediately find out the dimensionality of Newton's
constant to be $Nm^2/kg^2$, which translates to $[L]^3[M]^{-1}[T]^{-2}$.  
Using the fact that $[M]\sim [L]^{-1}$ and $[T]\sim [L]$ we see that the
mass dimensionality of $G$ is $[M]^{-2}$. So the power of the mass
dimension is negative for the coupling constant and the general
principles of renormalizability imply that the theory that we have, is
riddled with untamable divergences. This is one of the ways in which 
to look at the difficulty of quantizing gravity.

What does a dimensionful coupling mean, on top of unrenormalizability? If 
we look at tree-level perturbation theory, the amplitude for a 
scattering process is 
going to look something like $1+GE^2$ where the first term says that
the amplitude for nothing to change when there is no 
interaction is going to be 1: the probability for no scattering when 
there is no interaction is obviously a hundred percent. The second term 
comes from the tree-level interaction and should be proportional to the 
coupling constant $G$. To make that term dimensionally correct we 
should have an $E^2$, because $G\sim [M]^{-2}\sim [E]^{-2}$. Here $E$ is 
the characteristic energy of the process, by which we mean the center of 
mass energy of the particles in the scattering process, say. From this 
form of the total tree-level amplitude it is clear that for energies higher 
than $G^{-1/2}$, perturbation theory fails because the subleading term 
is of the same order of magnitude as the leading term. This energy scale 
at which gravity fails as a perturbative field theory is called the 
Planck Scale. 
Putting back the factors of $c$ and $\hbar$ by dimensional analysis, the 
Planck scale is $E_{Planck}=(\frac{\hbar c^5}{G})^{1/2} \sim 10^{19} GeV$. 
For energies below the Planck scale General Relativity is 
a useful theory but beyond that we need a theory which gives a better high 
energy (``ultraviolet") definition of gravity. A correct quantum theory of 
gravity is necessary to make sense of the processes at and above that 
scale.

Another problem with gravity is that in GR we have the freedom to make 
coordinate transformations. This is the gauge invariance of gravity. 
Usually the quantization of theories with gauge invariance is subtle, but 
not impossible. But in the case of gravity, the gauge freedom is the 
freedom for choice of coordinates, and this makes things doubly subtle. 
For instance, usually in canonical quantization in field theory, what we 
do 
is that we take a specific instant in time (``a spacelike hypersurface" if 
one wants to be pedantic) and then we define the canonical conjugates to 
the 
fields on that timeslice. The problem is that in gravity, coordinate 
freedom means that one's choice of time coordinate is not unique and we 
can move away from our choice of timeslice by doing a gauge 
transformation (a change of coordinates): 
but a gauge transformation is supposed to be an irrelevant transformation 
as far as the dynamics is considered, and we would not usually expect this 
sort of thing to happen. This issue is a manifestation of
what is often referred to as the 
Problem of Time in gravity\footnote{There is another problem associated 
with the choice of time: every time coordinate is defined in a specific 
coordinate system, and the question of whether there exists a consistent  
choice of time 
coordinate across different coordinate patches in spacetime is a 
non-trivial one.}. In fact there are a lot many other problems 
associated with quantizing gravity, both technical and conceptual. But 
the issue of non-renormalizability and the issue of the the meaning of 
the time coordinate are sufficient to give an idea that the standard 
procedures of 
quantization of field theories do not work when applied to gravity.
So it now becomes clear that its not such an unthinkable proposition to 
try to look for an indirect mechanism to incorporate gravity in the 
quantum scheme of things. And this is the relevance of string theory. From 
the next section on, I will start by defining string theory, and the 
reason for doing string theory will become clear when at the end of the 
day after we quantize string theory we will find a state 
in the string Hilbert space which can be interpreted as a graviton.

\section{Classical String Theory}

\subsection{Point particle}

Before starting with strings per se, I will start with the point particle 
moving freely in spacetime as some sort of motivation. For generality, 
spacetime will be assumed to be 
$D$-dimensional: $0,1,..., D-1$. We will find later on that the quantum 
consistency of string theory will put restrictions on the dimensionality 
of 
the spacetime in which the string propagates. So its essential that we 
do not fix the dimensionality before we begin\footnote{In my discussion 
of string theory, I will follow Polchinski very closely.}.

Free particles move along geodesics, which are obtained by minimizing the 
lengths of trajectories. The trajectory can be specified by giving the 
spatial position at time $X^0$, i.e, {\boldmath $X$}$=${\boldmath 
$X$}$(X^0)$. Here {\boldmath $X$}$=(X^{1},...,X^{D-1})$. But I want to 
make the formalism covariant and so I will introduce a parameter $\tau$ 
such that $X^0=X^0(\tau)$ and {\boldmath $X$}$=${\boldmath
$X$}$(\tau)$. In principle, one could obtain the original expression (i.e, 
{\boldmath $X$}$=${\boldmath $X$}$(X^0)$) by solving for $\tau$ in terms 
of 
$X^0$ and then plugging it into {\boldmath $X$}$=${\boldmath
$X$}$(\tau)$.

A covariant notation for the above trajectory would be 
$X^{\mu}=X^{\mu}(\tau)$. This can be thought of as a map (an embedding if 
one wants to be mathematically precise) of the real line (as denoted by 
$\tau$) into spacetime ($X^{\mu}$). So a one-dimensional line 
gets embedded in spacetime in some complicated way and we end up getting 
a curve, the worldline. Since I don't want the choice of parameter $\tau$ 
to 
affect my physics, I will decree that $X^{\prime\mu} (\tau ^{\prime} 
(\tau))=X^{\mu}(\tau)$, for any $\tau '=\tau '(\tau)$. This can be thought 
of as the freedom to do coordinate transformations (diffeomorphisms) on 
the 
worldline. These diffeomorphisms are the ``gauge freedom" in our theory. 

As I said, particles move along geodesics and these can be obtained by 
minimizing the proper time along the trajectory. (By trajectory, I mean 
worldline). So, if we choose the action for the system to be proportional 
to the proper time, the equation of motion obtained by varying the action 
would give us the geodesics. So lets take,
\beq
S_1=-m\int dT = -m\int \sqrt{-\eta_{\mu\nu}dX^{\mu}dX^{\nu}} 
=-m\int d\tau \sqrt{-\eta_{\mu\nu}\dot{X}^{\mu}\dot{X}^{\nu}},
\eeq 
where $\dot{X}^{\mu}=\frac{dX^{\mu}}{d\tau}$ and $dT$ is the element 
of proper time. Also,
\[\eta_{\mu\nu} = \left\{ \begin{array}{ll}
			    -1   & \mbox{if $\mu=\nu=0$}\\
			    +1   & \mbox{if $\mu=\nu\neq 0$}\\
			     0   & \mbox{otherwise.}
			   \end{array}
		  \right. \]		
The proportionality constant $-m$ is included so that the action reduces 
to the usual
\beq
\int (K.E.-P.E.)
\eeq
form in the non-relativistic limit.

Varying with respect to $X^{\mu}$ gives the Equations of Motion (EOM):
\begin{eqnarray}
\frac{d}{d\tau} \frac{\partial {\cal L}}{\partial \dot{X}^{\mu}}- 
\frac{\partial {\cal L}}{\partial X^{\mu}} = 0, \\
\Rightarrow m\frac{d}{d \tau}\Big( 
\frac{\dot{X}_{\mu}}{\sqrt{-\dot{X}_{\nu}\dot{X}^{\nu}}} \Big)=0.
\end{eqnarray}

Quantizing a theory is usually easier if we have a quadratic form for the 
action: essentially because we have good tools for handling linear 
equations of motion. In fact almost all the situations that we deal with in 
Quantum Mechanics or Quantum Field Theory start with a quadratic problem 
and then use perturbation theory as an approximation scheme to take care 
of the non-quadratic pieces. I write down the cases of the Simple Harmonic 
Oscillator and the free Klein-Gordon field to demonstrate that they are 
indeed purely quadratic: 
\begin{eqnarray}
S_{SHO}&=&\int dt \Big( \frac{1}{2} m\dot{x}^2 -\frac{1}{2} k x^2 \Big) \\
S_{KG}&=&\int d^{4}x \Big( \frac{1}{2} (\partial_{\mu} \varphi)^2 
-\frac{1}{2} m^2 {\varphi}^2 \Big).
\end{eqnarray}

Our worldline action on the other hand has an inconvenient square root and 
we can rewrite it 
in a convenient quadratic form by introducing a new variable $\eta$:
\beq
S_2=\frac{1}{2}\int d{\tau} \Big( 
\frac{\dot{X}_{\mu}\dot{X}^{\mu}}{\eta}-\eta m^2 \Big).
\eeq
I claim that the EOM from this action for the variable $\eta$ can be used to
rewrite this action, and the result is $S_1$. 
\beq
\delta_\eta S_2=0\Rightarrow \frac{1}{\eta^2} \dot{X}_{\mu}\dot{X}^{\mu} + 
m^2 =0.
\eeq
Solve for $\eta$ and plug back in $S_2$ and we get $S_1$. (Try it, its 
easy!). So we can say that $S_2$ is classically equivalent to $S_1$. 
Notice that $\eta$ is a non-dynamical variable, because the time 
derivative of $\eta$ does not appear in the action. It is described 
entirely in terms of the other variables in the theory (the $X^{\mu}$). In 
a field theory context, a variable of this form is referred to as an 
auxiliary field.

\subsection{Generalizing from the Point Particle to the String}
 
Now we want to write down an action for a string propagating in 
spacetime. The point-particle will serve as our toy-model and we want to use it as 
a hinge to make the generalization to the string. The first thing to note is that a point particle 
(a zero-dimensional object) sweeps out a line in spacetime (the 
worldline), whereas a string (which is a one-dimensional object) would 
sweep out a surface (a ``worldsheet"). Pushing our analogy further, as the 
action for the particle was proportional to the length of the worldline, 
the most natural action for the string is the area of the worldsheet. (1 
dimensional line : length $\Rightarrow$ 2-dimensional worldsheet : Area).
The area of a surface can be defined in terms of the metric on the 
surface as $\int d^2 x \sqrt {g}$. Here $g$ is the magnitude of the 
determinant of the metric. As a simple example one might consider 
Cartesian coordinates on the Euclidean plane: $x_1=x$, $x_2=y$. 
The 
metric 
tensor is $g_{ab}=\delta_{ab}$, i.e, $ds^2=dx^2+dy^2$. So $g=1$, and hence 
area $=\int dx dy \sqrt{1}=\int dx dy$ which is the expected formula. To give a slightly less trivial example, one can look at the same 
plane in polar coordinates: $x_1=r$, $x_2= \theta$. Now the non-zero 
elements of the metric are given by $g_{rr}=1$, and 
$g_{\theta\theta}=r^2$ (i.e, $ds^2=dr^2+r^2d\theta^2$). So, $g=r^2$. This 
means that the
area $=\int drd\theta \sqrt{r^2}=\int rdr d\theta$, again as expected.      

The moral of the above discussion is that to define an area for the 
worldsheet, we need a metric on it. Since we imagine that the worldsheet 
lives in spacetime, we can use the metric induced on the worldsheet due to 
the spacetime metric. What this means is that we use the idea of distance 
in spacetime to define a notion of distance on the worldsheet. Lets call 
the induced metric $h_{ab}$. Here $a$, $b$ are worldsheet indices, 
$\sigma^{a}=(\sigma^0, \sigma^1)=(\tau,\sigma).$ Then,
\begin{eqnarray}
h_{ab}d\sigma^{a}d\sigma^{b}&\equiv&\eta_{\mu\nu}dX^{\mu}dX^{\nu}|_{\Sigma} 
\\
&=&\eta_{\mu\nu}\Big(\frac{\partial X^{\mu}}{\partial 
{\sigma}^a}d{\sigma}^a\Big) \Big( \frac{\partial X^{\nu}}{\partial 
{\sigma}^b}d{\sigma}^b \Big) \\
&=&\big(\eta_{\mu\nu}\partial_{a} X^{\mu} \partial_{b} X^{\nu}\big) 
d\sigma^a d\sigma^b
\end{eqnarray}
So we see that $h_{ab}=\eta_{\mu\nu}\partial_{a} X^{\mu} \partial_{b} 
X^{\nu}=\partial_{a} X_{\mu} \partial_{b}X^{\mu}$.
In the first equality in the equations above, the restriction $|_{\Sigma}$ 
means that we are taking the spacetime 
distance element, but restricted to the surface of the worldsheet. So, in 
the next line, the variations in $X^{\mu}$, arise only through the changes 
in 
the worldsheet coordinates $\sigma^a$.  

Now, I can define the worldsheet action to be proportional to the area, 
with the area defined through the above metric.
\beq
S_{NG}=\frac{-1}{2\pi \alpha '}\int d\tau d\sigma \sqrt {-h}
\eeq 
with $h=\det(h_{ab})$, is called the Nambu-Goto action. The $-$ve sign on 
$h$ comes up because we choose the signature on the worldsheet as $(-,+)$. 
Here $\frac{-1}{2\pi \alpha '}$ is a proportionality constant. And $\alpha 
'$ is called the Regge slope because of certain obscure (and not 
so obscure) historical 
reasons.

Like in the point particle there is a coordinate freedom (often 
referred to as ``diff. invariance"):
\beq
X'^{\mu}(\tau ' (\tau, \sigma))=X^{\mu}(\tau, \sigma).
\eeq 
Note that these are two independent coordinate transformations, so there 
are two gauge degrees of freedom.

In the point particle we made the action into a quadratic form by 
introducing auxiliary fields (variables). We can do the same for the 
string. We introduce a metric on the worldsheet, $\gamma_{ab}$, which we 
treat as independent: not as derived from the spacetime metric. Then, let
\beq
S_{P}=\frac{-1}{4\pi \alpha '} \int d\tau d\sigma  
(-\gamma)^{1/2}\gamma^{ab}\partial_{a} X_{\mu} \partial_{b}X^{\mu}.
\eeq  
This is the so-called Polyakov action; the analogue of $S_2$ for the 
string. $\gamma=\det(\gamma_{ab})$. We can use the EOM for $\gamma_{ab}$ 
to reduce $S_P$ to $S_{NG}$. Lets do that. First of all,
\begin{eqnarray}
\delta(-\gamma)^{1/2}&=&\frac{1}{2(-\gamma)^{1/2}}\delta (-\gamma) \\
&=&\frac{(-\gamma)^{1/2}}{2}\frac{\delta \gamma}{\gamma}\\
&=&\frac{(-\gamma)^{1/2}}{2}\gamma^{ab}\delta \gamma_{ab} \\
&=&-\frac{(-\gamma)^{1/2}}{2}\gamma_{ab}\delta \gamma^{ab}.
\end{eqnarray}
where I have used the fact that $\gamma^{ab}\gamma_{ab}=\delta_b^b=2$ and 
so, $\gamma^{ab}\delta \gamma_{ab}+\gamma_{ab}\delta \gamma^{ab}=0$. Thus 
$\delta_{\gamma_{ab}}S_P=0$ 
means,
\begin{eqnarray}
0&=&\frac{-1}{4\pi \alpha '}\int d\tau d\sigma
(-\gamma)^{1/2}\delta \gamma^{ab}\Big( \partial_{a} X_{\mu} \partial_{b}X^{\mu}-\frac{1}{2}\gamma_{ab}(\gamma^{cd}\partial_{c} X_{\mu} 
\partial_{d}X^{\mu})\Big)\\
&=&\frac{-1}{4\pi \alpha '}\int d\tau d\sigma
(-\gamma)^{1/2}\delta \gamma^{ab}\Big( h_{ab}-\frac{1}{2}\gamma_{ab}(\gamma^{cd}h_{cd})\Big)
\end{eqnarray} 
This implies that 
\beq
h_{ab}=\frac{1}{2}\gamma_{ab}(\gamma^{cd}h_{cd}),
\eeq
and thus,
\beq
h=\det(h_{ab})=\frac{1}{4} 
(\gamma^{cd}h_{cd})^2\det(\gamma_{ab})=\frac{\gamma}{4}(\gamma^{cd}h_{cd})^2.	
\eeq
Using this in the expression for $S_P$, we find,
\begin{eqnarray}
S_P&=&\frac{-1}{4\pi \alpha '}\int d\tau d\sigma(-\gamma)^{1/2}\gamma^{ab} 
h_{ab}\\
&=&\frac{-1}{4\pi \alpha '}\int d\tau d\sigma \big( 2(-h)^{1/2} \big)\\
&=&\frac{-1}{2\pi \alpha '}\int d\tau d\sigma(-h)^{1/2}\\
&=&S_{NG}
\end{eqnarray}

On top of the ``diff-invariance" of $S_{NG}$, $S_P$ has an invariance 
under $\gamma_{ab} \rightarrow e^{2\omega(\tau, \sigma)}\gamma_{ab}$, 
because the combination $(-\gamma)^{1/2}\gamma^{ab}$ is invariant under 
this. This is called a Weyl transformation. So the Polyakov action has 
three gauge symmetries: the two diffs, and the Weyl.

Because of its quadratic structure, the Polyakov action is more suitable 
for quantization, especially in the path-integral formalism.

\subsection{The String EOM from $S_P$}

EOM for $\gamma_{ab}$ comes from $\delta_{\gamma_{ab}}S_P=0$. This implies 
that,
\begin{eqnarray}
T_{ab}&\equiv&\frac{4\pi}{(-\gamma)^{1/2}}\frac{\delta S_P}{\delta 
\gamma^{ab}}\\
&=&\frac{-1}{\alpha '}\Big( h_{ab} -\frac{1}{2} \gamma_{ab} (\gamma^{cd}h_{cd}) \Big) \\
&=&0.
\end{eqnarray}

EOM for $X^{\mu}$ arises from,
\begin{eqnarray}
0&=&\delta_X S_P \\
&=&\frac{-1}{4\pi \alpha '}\int d\tau d\sigma (-\gamma)^{1/2} \gamma^{ab} 
\Big[ \partial_a (\delta X^{\mu}) \partial_b X_{\mu} + \partial_a 
X^{\mu}\partial_b(\delta X_{\mu}) \Big] \\
&=&\frac{-1}{2\pi \alpha '} \int d\tau d\sigma (-\gamma)^{1/2} 
\gamma^{ab}\partial_a
X^{\mu}\partial_b(\delta X_{\mu})\\
&=&\frac{1}{2\pi \alpha '}\int d\tau d\sigma\partial_b \Big( 
(-\gamma)^{1/2}
\gamma^{ab}\partial_a
X^{\mu} \Big) \delta X^{\mu}.
\end{eqnarray}
In the third line I have used the symmetry of $a$ and $b$ and in the last 
line I have done an integration by parts. To justify this integration by 
parts, I have to make sure that the boundary term vanishes at the spatial 
and timelike boundaries ($\sigma$ an $\tau$ respectively). For timelike 
boundaries the vanishing of the boundary term is true by assumption 
involved in the variational principle. For the spatial case we can 
avoid the issue altogether by imposing periodicity: $X^{\mu}(\tau, 
\sigma)=X^{\mu}(\tau, \sigma+\ell)$ for some $\ell$ and thus by declaring 
that the string is a closed string with no boundaries. Since 
$\delta 
X^{\mu}$ is arbitrary this means that the EOM is, \begin{eqnarray}
\partial_b \Big(
(-\gamma)^{1/2}
\gamma^{ab}\partial_a
X^{\mu} \Big) &=&0 
\end{eqnarray}
or, which is the same thing,
\begin{eqnarray}
\nabla ^2 X^{\mu}=0
\end{eqnarray}
where $\nabla$ stands for the covariant derivative. So $X^{\mu}$ 
satisfies the wave equation.

\section{Quantization of the String}

The string that we have looked at is what is called the bosonic closed 
string. Closed, because the boundary conditions that we imposed 
on the string at the end of the last section are closed string boundary 
conditions. And bosonic, because none of the fields in the worldsheet 
theory are 
fermionic. We will quantize them using commutators, not anticommutators as  
is always done for bosons.
Also, when we quantize the string, we will 
find that the states in the Hilbert space are all bosonic from the 
spacetime 
perspective as well. 

I will use a variation of the so-called light-cone gauge to quantize the 
string. This is an ugly approach, but its the fastest way to get to the 
spectrum of states of the string. Also, I want to do string theory 
with the bare minimum of field theory tools. There 
are more elegant, powerful, and 
general methods for quantizing the string: starting with 
Polaykov's path integral technique, Conformal Field Theory and the BRST 
formalism. But since these lectures are only an invitation, the reader 
should look elsewhere for an introduction to these very 
useful tools. 

In any theory with gauge degrees of freedom we have a choice: we can 
quantize before fixing the gauge or fix the gauge and then quantize. 
Quantizing Electrodynamics in the Gupta-Bleuler formalism is an example of 
the former. But quantizing it in the Coulomb gauge (like Fermi did) is an 
example of the latter. Light-cone gauge quantization is an example of 
a gauge-fixed scheme for the string: we fix the three gauge redundancies 
associated with the diff, Weyl invariances, and then quantize.   

First define,
\begin{eqnarray}
X^{\pm}&=&\frac{1}{\sqrt 2}(X^0 \pm X^1)\\
X^{i}&=&X^2,...,X^{D-1},
\end{eqnarray}
with
\begin{eqnarray}
X^{\mu}X_{\mu}&=&-(X^0)^2+(X^1)^2+...+(X^{D-1})^2 \\
&=&-X^{+}X^{-}-X^{-}X^{+}+(X^2)^2+...+(X^{D-1})^2\\
\Rightarrow X^{\mu}X_{\mu}&=&-X^{+}X^{-}-X^{-}X^{+}+{X^{i}}^2.
\end{eqnarray}
So, 
\begin{eqnarray}
\eta^{+-}=\eta^{-+}=-1,\eta^{ii}=+1,
\end{eqnarray}
with rest of the elements of $\eta$ equal to zero. Of course, in the above 
definition, there is no summation on $i$.

There are three gauge invariances:  2 co-ord transformations (diff) and 
one Weyl. So we are free to impose three conditions. I will take them to 
be,
\begin{eqnarray}
X^{+}&=&\tau\\
\partial_\sigma \gamma_{\sigma\sigma}&=&0\\
\det(\gamma_{ab})&=&-1.
\end{eqnarray}
This gauge choice will be our light-cone gauge. Note that among other 
things this choice implies that $\gamma_{\sigma\sigma}$ is a 
function of only $\tau$. Now lets look 
at the Polyakov action in this gauge. Before that I have to see what the 
worldsheet metric and its inverse are in this gauge. First note that the 
inverse of a $(2\times 2)$ matrix $M$ is,
\[ M^{-1} = \left( \begin{array}{c}
+a \ \ +b 	\\
+c \ \ +d  \end{array} \right)^{-1}  = \frac{1}{\det M}\left( 
\begin{array}{c}
+d \ \ -b         \\
-c \ \ +a  \end{array} \right) 
\]
So \[ \gamma^{ab}=\left( \begin{array}{c}
\gamma^{\tau\tau} \ \ \gamma^{\tau\sigma}        \\
\gamma^{\tau\sigma} \ \ \gamma^{\sigma\sigma}  \end{array} \right) =  
\frac{1}{\det (\gamma_{ab})} \left(
\begin{array}{c}
\gamma_{\sigma\sigma} \ \ -\gamma_{\tau\sigma}        \\
-\gamma_{\tau\sigma} \ \ \gamma_{\tau\tau}  \end{array} \right) \\
=\left(\begin{array}{c}
-\gamma_{\sigma\sigma}(\tau) \ \ \ \gamma_{\tau\sigma}(\tau, \sigma) \\
\gamma_{\tau\sigma}(\tau, \sigma) \ \ \  
\gamma_{\sigma\sigma}^{-1}(1-\gamma_{\tau\sigma}^2) \end{array} \right)
\]
where I have used the fact that $\det(\gamma_{ab})=-1$. Writing this 
equation out explicitly, we see that 
$\gamma_{\tau\tau}\gamma_{\sigma\sigma}-\gamma_{\tau\sigma}^2=-1$, which 
has been used to solve for $\gamma_{\tau\tau}$ in the last step. Using 
these, the Polyakov Lagrangian becomes, 
\begin{eqnarray}
L_P&=&\frac{-1}{4 \pi \alpha '}\int_0^{\ell} d\sigma \gamma^{ab} 
\partial_a X^{\mu} \partial_b X_{\mu} \\
&=&\frac{-1}{4 \pi \alpha '}\int_0^{\ell} d\sigma \Big[ 
-\gamma_{\sigma\sigma} (\partial_{\tau}X^{\mu}\partial_{\tau}X_{\mu} + 2 
\gamma_{\tau\sigma} \partial_{\tau}X^{\mu} \partial_{\sigma} 
X_{\mu}+\gamma_{\sigma\sigma}^{-1}(1-\gamma_{\tau\sigma}^2) 
\partial_{\sigma}X^{\mu}\partial_{\sigma}X_{\mu} \Big] \\
&=&\frac{-1}{4 \pi \alpha '}\int_0^{\ell} d\sigma 
\begin{array}{c}
\Big[ \gamma_{\sigma\sigma}(2\partial_{\tau}X^{-}-
\partial_{\tau}X^{i}\partial_{\tau}X^{i}) + 
2\gamma_{\tau\sigma}(-\partial_{\sigma}X^{-}+\partial_{\tau}X^{i}
\partial_{\sigma}X^{i})+ \\
+\gamma_{\sigma\sigma}^{-1}(1-\gamma_{\tau\sigma}^2)
\partial_{\sigma}X^{i}\partial_{\sigma}X^{i} \Big]
\end{array} 
\end{eqnarray}
This is turn implies that
\[ 
L_P=\frac{-1}{4 \pi \alpha '}.\ 2 
\gamma_{\sigma\sigma}\partial_{\tau}\Big(\int_0^\ell d\sigma 
X^{-}(\tau,\sigma)\Big) + 
\]
\[
+\frac{-1}{4 \pi \alpha '}\int_0^{\ell} d\sigma
\begin{array}{c}
\Big[ 
-\gamma_{\sigma\sigma}\partial_{\tau}X^{i}\partial_{\tau}X^{i}-2
\gamma_{\tau\sigma}
(\partial_{\sigma}X^{-}-\partial_{\tau}X^{i}
\partial_{\sigma}X^{i})
+\gamma_{\sigma\sigma}^{-1}(1-\gamma_{\tau\sigma}^2)
\partial_{\sigma}X^{i}\partial_{\sigma}X^{i} \Big].
\end{array} \]
Defining 
\begin{eqnarray}
x^{-}(\tau)&\equiv&\frac{1}{\ell}\int_0^{\ell}d\sigma 
X^{-}(\tau,\sigma) \\
Y^{-}(\tau,\sigma)&\equiv&X^{-}(\tau,\sigma)-x^{-}(\tau),
\end{eqnarray}
this reduces to,
\[ L_P=\frac{-\ell}{2 \pi \alpha '}\gamma_{\sigma\sigma}\partial_\tau 
x^{-} + \]
\vspace{-7 mm}
\beq
-\frac{1}{4 \pi \alpha '}\int_0^{\ell} d\sigma
\Big[
-\gamma_{\sigma\sigma}\partial_{\tau}X^{i}\partial_{\tau}X^{i}-2
\gamma_{\tau\sigma}
(\partial_{\sigma}Y^{-}-\partial_{\tau}X^{i}
\partial_{\sigma}X^{i})
+\gamma_{\sigma\sigma}^{-1}(1-\gamma_{\tau\sigma}^2)
\partial_{\sigma}X^{i}\partial_{\sigma}X^{i} \Big].
\eeq
In the last line I have used the fact that 
$\partial_{\sigma}Y^{-}=\partial_{\sigma}X^{-}$.

Now, we have our action. First, note that $Y^{-}$ does not appear with 
time derivatives, so is an auxiliary field. Vary with respect to 
$Y^{-}$. The resulting EOM\footnote{its not really an EOM 
because it will not involve time derivatives, but we will call it by that 
name anyway.} is $\partial_\sigma \gamma_{\tau\sigma}=0$. (Another way to 
look at this equation is to think of $Y^{-}$ as a Lagrange multiplier 
that constrains $\gamma_{\tau\sigma}$ to satisfy this equation.).	

So far everything I have done is generic to the bosonic string. Now I am 
going to specialize to the closed bosonic string. In this case, we have 
some more gauge freedom that has not been fixed by the light-cone 
conditions. This is because for any value of $\tau$, we can choose 
the $\sigma=0$ point on the string arbitrarily. We can fix this 
arbitrariness almost fully by stipulating that $\hat e_\tau$ (the unit 
tangent vector along $\sigma=0$ at some point $(\tau,\sigma=0)$) is 
orthogonal to $\hat e_{\sigma}$ at the same point. That is, 
$\gamma_{ab}(\hat e_\tau,\hat e_{\sigma=0})=0$. In other words, 
$\gamma_{\tau\sigma}(\tau,0)=0$.  

Now, once we fix this, $\gamma_{\tau\sigma}(\tau,0)=0$ and 
$\partial_\sigma \gamma_{\tau\sigma}=0$ (obtained by varying $Y^{-}$) 
together imply that $\gamma_{\tau\sigma}=0$ everywhere. Putting this in 
the expression for the Lagrangian, we get
\beq
\label{finalp}
L_P=\frac{-\ell}{2 \pi \alpha '}\gamma_{\sigma\sigma}\partial_\tau
x^{-} + \frac{1}{4 \pi \alpha '}\int_0^{\ell} d\sigma(\gamma_{\sigma\sigma}
\partial_{\tau}X^{i}\partial_{\tau}X_{i}-\gamma_{\sigma\sigma}^{-1}\partial_{\sigma}X^{i}\partial_{\sigma}X^{i}).
\eeq
Now we can do the usual things that one does in the classical canonical 
formalism. Thus,
\begin{eqnarray}
p_{-}&=&\frac{\partial L_P}{\partial(\partial_\tau x^{-})}=-\frac{\ell}{2 
\pi \alpha '}\gamma_{\sigma\sigma} \\
p_{-}&=&\eta_{-+}p^{+}=-p^{+} \Rightarrow p^{+}=\frac{\ell}{2
\pi \alpha '}\gamma_{\sigma\sigma} \\
\Pi^{i}&=&\frac{\delta L_P}{\delta(\partial_\tau X^{i})}=\frac{1}{2
\pi \alpha '}\gamma_{\sigma\sigma} \partial_\tau 
X^{i}=\frac{p^{+}}{\ell}\partial_\tau X^{i}
\end{eqnarray}
We can easily use these expressions to solve for $\partial_\tau X^{i}$ in 
terms 
of $\Pi^{i}$. Then the canonical Hamiltonian defined as the usual 
Legendre transform becomes,
\begin{eqnarray}
H_P&=&p_{-}\partial_\tau x^{-}+\int_0^{\ell}d \sigma \Pi^{i} \partial_\tau 
X^{i} - L_P \\
&=&\frac{\ell}{2p^{+}}\int_0^{\ell}d \sigma \Pi^{i} 
\Pi^{i}+\frac{\ell}{2(2\pi\alpha ')^2 p^{+}}\int_0^{\ell}d \sigma 
\partial_\sigma X^{i} \partial_\sigma X^{i}.
\end{eqnarray}
The EOM for $X^{i}$ can be obtained by varying $L_P$ with respect to 
$X^{i}$. The Euler-Lagrange equation looks like:
\begin{eqnarray}
\partial_{\sigma}\Big(\frac{\delta L_P}{\delta(\partial_\sigma 
X^{i})}\Big)+\partial_\tau\Big(\frac{\delta L_P}{\delta(\partial_\sigma
X^{i})}\Big)-\frac{\delta L_P}{\delta X^{i}}&=&0 \\
\Rightarrow\partial_{\sigma}(-2\gamma_{\sigma\sigma}^{-1} 
\partial_{\sigma} X^{i}) +
\partial_\tau (2\gamma_{\sigma\sigma}\partial_{\tau}X^{i})&=&0 .
\end{eqnarray}
We know from the gauge-fixing condition that $\partial_\sigma 
\gamma_{\sigma\sigma}=0$. And,
\begin{eqnarray}
\partial_{\tau}\gamma_{\sigma\sigma}&=&\frac{-2\pi\alpha 
'}{\ell}\partial_{\tau} p_{-}=\frac{-2\pi\alpha
'}{\ell}\Big(-\frac{\partial H}{\partial x^{-}}\Big) =0.
\end{eqnarray}
The second equality is nothing but a Hamilton's equation of motion. The 
last equality on the other 
hand is a direct 
consequence of the explicit form of the Hamiltonian $H_P$ that we wrote 
down earlier: $H_P$ is independent of $x^{-}$. From these, we conclude 
that $\gamma_{\sigma\sigma}$ is independent of both $\sigma$ and $\tau$. 
So 
the EOM becomes, 
\begin{eqnarray}
\partial_\tau^2 X^{i}=\gamma_{\sigma\sigma}^{-2}\partial_\sigma^2 X^{i}.
\end{eqnarray} 
Defining the constant $\gamma_{\sigma\sigma}^{-1}\equiv c$ (unrelated to 
the velocity of light), we have the 
wave 
equation:
\beq
(\partial_\tau^2-c^2\partial_\sigma^2)X^{i}=0.
\eeq 

From our experience with quantizing free field theories (e.g., Free 
Klein-Gordon field), we expect that the physical interpretation of the 
quantization is going to be transparent if we do a mode expansion. And 
what is the natural mode expansion in the case of the string? We saw that 
the fields in the string worldsheet satisfy the wave equation. We also 
know from the closed string boundary condition that $X^{i}(\tau, 
\sigma)=X^{i}(\tau,\sigma+\ell)$. Periodicity means that the function 
can be expanded in a Fourier series in that coordinate. So lets do that! 
Let
\begin{eqnarray}
\label{mode}
X^{i}&=&\sum_{n=-\infty}^{\infty} f^{i}_n(\tau)\exp(\frac{2n\pi i 
\sigma}{\ell}).
\end{eqnarray}
Plugging this form into the wave equation above, we get,
\begin{eqnarray}
f^{i\prime\prime}_n(\tau)-c^2\Big( \frac{2n\pi i }{\ell}\Big)^2 
f^{i}_n(\tau)=0.
\end{eqnarray}
where I use a prime to denote differentiation with respect to $\tau$.
Note that for $n=0$, this means that we have 
$f^i_0(\tau)=\frac{p^{i}}{p^{+}}\tau+x^i$, whereas for $n\neq 0$, we have
a Simple Harmonic Oscillator equation which can be solved as,
\beq
f^{i}_n(\tau)=i\Big(\frac{\alpha 
'}{2}\Big)^{1/2}\Big[-\frac{\alpha^i_{-n}}{n}\exp\Big(\frac{2ni\pi 
c\tau}{\ell}\Big)+\frac{\tilde\alpha^i_n}{n}\exp\Big(\frac{-2ni\pi 
c\tau}{\ell}\Big)\Big].
\eeq
In the above expressions, the new quantities that have been introduced are 
integration constants, and I have taken them in these specific forms so 
that there is maximum agreement with the notation in the literature. The 
untilded modes 
are called the ``left-movers" and the tilded ones, the ``right-movers".
Now we can put these expressions back in (\ref{mode}), and we get after 
some relabeling of summation variable $n$,
\beq
X^{i}(\tau,\sigma)=x^{i}+\frac{p^{i}}{p^{+}}\tau+i\Big(\frac{\alpha
'}{2}\Big)^{1/2}\sum_{n\neq 
0}\Big[\frac{\alpha^i_{n}}{n}\exp\Big(\frac{-2ni\pi(\sigma+c\tau)}{\ell}
\Big)+\frac{\tilde\alpha^i_n}{n}\exp\Big(\frac{ 
2ni\pi(\sigma-c\tau)}{\ell}\Big)\Big]
\eeq 
This is the closed string mode expansion which will be of great use to us 
when we quantize the theory next. 

The variables in the gauge-fixed theory are $x^{-}$ and $X^{i}$. So to
quantize the theory in the canonical prescription, we have to impose
canonical commutators between these variables and their corresponding
canonical momenta (which we calculated in the Hamiltonian formalism).
\begin{eqnarray} 
[x^{-},p_{-}]=i \Rightarrow [x^{-},p^{+}]=-i
\end{eqnarray} 
\vspace{-13mm} 
\begin{eqnarray} 
\label{dirac}
[X^{i}(\sigma),\Pi^{j}(\sigma^{\prime})]=i\delta^{ij}
\delta(\sigma-\sigma^{\prime}). 
\end{eqnarray}
We want to express these commutators as commutators of the left- and  
right-moving modes.
Plugging in the mode expansion and going through the kind of 
calculations familiar from free field theory, we get the mode algebra:
\begin{eqnarray}
[x^i,p^j]=i\delta^{ij}
\end{eqnarray}
\vspace{-13mm}
\begin{eqnarray}
[\alpha^i_{n},\alpha^j_{m}]=m\delta^{ij}\delta_{m,-n}
\end{eqnarray}
\vspace{-13mm}
\begin{eqnarray}
[\tilde\alpha^i_{n},\tilde\alpha^j_{m}]=m\delta^{ij}\delta_{m,-n}
\end{eqnarray}
These are the non-vanishing commutators between the modes. Its easy to 
note 
from the above algebra that the modes satisfy a sort of a scaled version 
of 
the Harmonic Oscillator algebra and that the 
$\frac{\alpha^{i}_m}{\sqrt{|m|}}$ are like annihilation operators for $m 
> 0$ 
and like creation operators for $m < 0$. Similar statements hold true for 
the left-moving modes. So we see that the algebra that determines the 
states in the Hilbert space is a combination of Heisenberg algebras 
($[x,p]=i$ kind of thing), and harmonic oscillator type mode algebras. So 
we can construct the Hilbert space by starting with a ground state defined 
by
\begin{eqnarray}
p^{+}|0,0;k\rangle =k^{+}|0,0;k\rangle,\ \ p^{i}|0,0;k\rangle 
=k^{i}|0,0;k\rangle 
\end{eqnarray}
\vspace{-13mm}
\begin{eqnarray}
\alpha^i_{n}|0,0;k\rangle=\tilde\alpha^i_{n}|0,0;k\rangle =0 \ \ \forall \
n>0.
\end{eqnarray}
We can act on these ground states with the creation operators and 
construct a Fock space 
analogous to the 
one familiar from field theory. So, upto normalization, the general state 
would look like 
\begin{eqnarray}
\label{state}
|N,\tilde N; 
k\rangle=\prod_{i=2}^{D-1}\prod_{n=1}^{\infty}(\alpha^i_{-n})^{N_{in}}(\tilde 
\alpha^i_{-n})^{\tilde N_{in}}|0,0;k\rangle,
\end{eqnarray}
which is a straightforward generalization of the harmonic oscillator Fock 
space. We can define a quantity called the level,
\beq
N=\sum_{i=2}^{D-2}\sum_{n=1}^{\infty}n N_{in},
\eeq
for the left-movers and and an exactly analogous quantity on the 
right-mover side, $\tilde N$. In my definition of the general state above, 
I have characterized it by specifying the levels. Actually, the states, as 
I have 
written them 
down in (\ref{state}), are a little more general than what is actually 
allowed: in fact, only states with $N=\tilde N$ are allowed in the 
Hilbert space of the string. So, a general state is of the form 
(\ref{state}), with the extra condition that $N=\tilde N$, which is called 
the level-matching condition.  

The level-matching condition can be derived using the idea that on the 
closed string there is a freedom to do (rigid) translations along 
$\sigma$. This symmetry\footnote{I fixed almost all of this symmetry when 
I fixed 
$\gamma_{\tau\sigma}(\tau,0)=0$ above eqn.(\ref{finalp}); but not all. 
The idea is that I initially had the freedom to choose the origin of 
$\sigma$ independently for any value of $\tau$. But once I fix 
$\gamma_{\tau\sigma}(\tau,0)=0$, what remains is only the freedom to 
do a 
rotation of the origin ($\sigma=0$) that is the same for all values of 
$\tau$: 
this symmetry is what is being taken care of 
now.} is generated\footnote{This will be discussed in the Notes 
at the end of this section.} 
quantum mechanically by an operator 
proportional to $N-\tilde N$. Since its a symmetry, the states have to be 
invariant under the symmetry operation, which in turn means that they 
have to be annihilated by the generator. Thus $N-\tilde N$ must 
annihilate the states and so we get the condition that $N=\tilde N$ for 
the acceptable states in the Hilbert space.

Now we turn to writing the Hamiltonian $H_P$ in terms of the modes. Direct 
substitution in terms of the mode expansion gives after some algebra, 
\begin{eqnarray}
H_P&=&\frac{p^{i}p^{i}}{2p^{+}}+\frac{1}{2\alpha' p^{+}}\sum_{n\neq 
0}(\alpha^i_{-n}\alpha^i_n+\tilde\alpha^i_{-n}\tilde\alpha^i_n) \\
&=&\frac{p^{i}p^{i}}{2p^{+}}+\frac{1}{2\alpha' 
p^{+}}\sum_{n=1}^{\infty} 
\big[ 2(\alpha^i_{-n}\alpha^i_n+\tilde\alpha^i_{-n}
\tilde\alpha^i_n)+\delta^{ii}n+\delta^{ii}n \big] \\
&=&\frac{p^{i}p^{i}}{2p^{+}}+\frac{1}{\alpha'
p^{+}}\Big[ \sum_{n=1}^{\infty} 
(\alpha^i_{-n}\alpha^i_n+\tilde\alpha^i_{-n}
\tilde\alpha^i_n) + \frac{D-2}{2}\sum_{n=1}^{\infty}n+
\frac{D-2}{2}\sum_{n=1}^{\infty}n \Big]\\
&=&\frac{p^{i}p^{i}}{2p^{+}}+\frac{1}{\alpha'
p^{+}}\Big[ \sum_{n=1}^{\infty}
(\alpha^i_{-n}\alpha^i_n+\tilde\alpha^i_{-n}
\tilde\alpha^i_n) + (D-2)\sum_{n=1}^{\infty}n \Big]
\end{eqnarray}
In the second line I have used the algebra of the modes to rewrite the 
$n<0$ terms in a normal ordered (i.e, creation operators to the left of 
annihilation operators) way. In the last lines I have explicitly written 
the trace over the transverse directions ($i=2$ to $D-1$) of $\delta^{ii}$ 
as $D-2$. 

The problem with this sum is that it contains divergent pieces. I will 
take care of this divergence in what might appear as a cavalier way, but 
which 
can actually be justified using the principles of renormalization. I will 
start by noting that $\sum_{n=1}^{\infty}n = \zeta(-1)$, where $\zeta(s)$ 
is 
the Riemann Zeta function. The original definition of the Zeta function 
($\zeta(s)=\sum_{n=1}^{\infty}1/n^s$) 
divergences on the left 
half-plane, but it has a unique analytic continuation there, and we will 
take the value of the above divergent sum to be the value of the analytic 
continuation. It is known from the analytic continuation that 
$\zeta(-1)=-1/12$. So we end up with the rather curious formula 
$\sum_{n=1}^{\infty}n=-1/12$. I will discuss this further in the Notes at 
the end  
of this section. 

The next thing to note is that the operator $\sum_{n=1}^{\infty}
(\alpha^i_{-n}\alpha^i_n+\tilde\alpha^i_{-n}
\tilde\alpha^i_n)$, when acting on a state gives the total 
level ($N+\tilde N$) of that state. This is easily checked for low-lying 
states and one can convince oneself that the pattern obviously 
generalizes. For example, by repeated use of the oscillator mode algebra 
to 
bring the annihilation operators all the way to the right (where they act 
on the vacuum and annihilate it), one can show that
 \[ \Big[ \sum_{n=1}^{\infty}
(\alpha^i_{-n}\alpha^i_n+\tilde\alpha^i_{-n}
\tilde\alpha^i_n)\Big] \alpha^{i}_{-2}\tilde 
\alpha^j_{-2}|0,0;k\rangle=\ 4 \ \alpha^{i}_{-2}\tilde 
\alpha^j_{-2}|0,0;k\rangle
\]
Using all this in the expression for the Hamiltonian, we get
\begin{eqnarray}
\label{ham}
H_P&=&\frac{p^{i}p^{i}}{2p^{+}}+\frac{1}{\alpha'
p^{+}}\Big[ N+\tilde N -\frac {(D-2)}{12} \Big]
\end{eqnarray}
Now I want to motivate that $p^{-}=H_P$. First, look at 
$p^{\mu}x_{\mu}=-Et+{\bf p}.{\bf x}$. Notice that the conjugate variable 
to $t$ is $E$, the energy. Now if we write the same $p^{\mu}x_{\mu}$ in 
light-cone coordinates, it looks like $-p^{-}X^{+}-p^{+}X^{-}+p^i X^{i}$.
Remember that our definition of light-cone gauge choice included 
setting $X^{+}=\tau$, a timelike variable. So, the conjugate quantity to 
$X^{+}$ should be an energy, the Hamiltonian. So $p^{-}=H_P$, which is 
what I set out to motivate. This argument can actually me made more 
rigorous, but we will not try to do so. We will take this as sufficient 
justification for concluding that $p^{-}=H_P$.

Using this, and the expression (\ref{ham}) that we found for the 
Hamiltonian, we come 
to the conclusion that the mass-squared operator, defined as 
$m^2=p^{\mu}p_{\mu}=2p^{+}p^{-}-p^ip^i$ in the light-cone gauge, is
\begin{eqnarray}
m^2&=&\frac{2}{\alpha'}\Big(N+\tilde N+\frac{2-D}{12}\Big).
\end{eqnarray}

All this effort in going through the Hamiltonian etc. and writing down an
expression for the mass operator was so that we could calculate the masses
of the states in the Hilbert space. We will calculate the masses for the
first few levels. First lets look at the ground state or the vacuum, 
$|0,0;k\rangle$. 
Here, $N=\tilde N=0$, so $m^2=(2-D)/6\alpha '$. From everyday 
experience we know that our spacetime has at least 4 dimensions: there 
could be more that are small in size and are experimentally 
inaccessible, but definitely there cannot be less than 4 spacetime 
dimensions. But we see from the above expression for the mass-squared of 
the 
ground-state that it is forced to be negative if the number of spacetime 
dimensions is greater than two. Such a state is called a tachyon and we see 
that the bosonic closed string has a tachyonic vacuum. This is not a good 
thing, but the good news is that in more realistic string theories, the 
so-called Superstring theories, there is no tachyon. We can think of the 
bosonic string as a toy model. 

Now, lets press on and see what the higher states look like. Because of 
the level-matching restriction, the first 
excited state is $\alpha^i_{-1}\tilde \alpha^j_{-1}|0,0;k\rangle$. Here 
$i$ and $j$ run from 2 to $D-1$. So there are $(D-2)^2$ states at this 
level. 
They  
have $m^2=(26-D)/6\alpha '$. Its known from the representation theory of 
the 
Lorentz group that these $(D-2)^2$ states can form full representations of 
the $D$-dimensional Lorentz group only if these states are 
massless\footnote{Remember that in 4 dimensions a massless $A^{\mu}$-field 
can form a full vector representation with (4-2)=2 
components, the two polarizations of the photon. The D-dimensional tensor 
analogue of this, is what we are referring to.}. This would imply that 
$D=26$. So we see that for Lorentz invariance to be preserved at the 
quantum level, we have to have 26 dimensions for the bosonic string. 
This is called the critical dimension. For superstring theories an 
analogous calculation would show that the critical dimension is 10. 
In anycase, we have a 
massless 2nd rank tensor representation of the Lorentz group as a state in 
the Hilbert space of string theory. The symmetric part of a massless 
second rank tensor is what we call a graviton: we saw that in one of 
the the earlier sections. So the string Hilbert space contains a graviton! 
In the realistic superstring theories, we will manage to get rid of the 
tachyon, but we will still have the graviton. In fact, all closed string 
theories contain a graviton state.

\subsection{Notes}

This subsection is meant to smooth out the few rough spots in our 
quantization of the closed bosonic string. I will be brief and will not 
develop everything systematically because these are not central issues.

First, I will clarify how $N-\tilde N$ becomes proportional to the 
generator of 
$\sigma$-translations. We needed that to demonstrate the level-matching 
condition earlier. The way to see this is to note that by 
a generator of $\sigma$ translations we mean an operator $P$ such that
$[P,X^i(\sigma)]=i\partial_\sigma X^i(\sigma)$. (Symmetries and their 
generators are discussed in many 
books on QM and QFT. An especially nice place to look 
this up 
would be ``Conformal Field Theory", by Di Francesco et al.). Due to the 
commutation relations (\ref{dirac}), it can be checked that the operator 
defined by 
\beq
P=-\int_0^{\ell}d\sigma \Pi^i\partial_\sigma X^i
\eeq
satisfies this relation. So this is the operator that generates $\sigma$ 
translations. Now, plug in the mode expansion for $X^i$ and $\Pi^i\sim 
\partial_\tau X^i$ in this expression for $P$. After the dust settles, we 
will find that 
\beq
P=-\frac{2\pi}{\ell}\sum_{n=1}^{\infty}(\alpha^i_{-n}\alpha^i_{n}
-\tilde\alpha^i_{-n}\tilde\alpha^i_{n})=-\frac{2\pi}{\ell}(N-\tilde N).
\eeq
So indeed, the generator of $\sigma$ translations is proportional to 
$N-\tilde N$ as promised.

Another point that could probably use some discussion is the result
$\sum_{n=1}^{\infty}n=-1/12$. The basic reason why its alright to pull the 
Zeta-function stunt (``regularization" would be the more conventional term 
in the literature) is because what we are doing there is equivalent to a  
summation with an $\epsilon$ cutoff: $\sum n\exp(-n\epsilon)$. This sum 
can 
be done (Try it, its not too hard!!), 
and if one writes the result as a power series in $\epsilon$, there will 
be a single pole term, an $\epsilon$-independent term equal to $-1/12$ and 
then higher powers in 
$\epsilon$. As we take the limit $\epsilon\rightarrow 0$, the only 
relevant terms are the first two. Now, the divergent $\epsilon$-dependent 
first term can be canceled by the addition of a counterterm to the 
original Polyakov action: in fact, since $\epsilon$-dependence is a scale 
dependence which signifies the violation of Weyl invariance, we must add 
this 
counterterm in order to preserve the Weyl symmetry. This is the rationale 
for 
neglecting the divergent $\epsilon$-pole divergence and taking the sum to 
be just the $-1/12$.

\section{Prospects}

We have quantized the string and found that there is a graviton. But 
this is just the beginning. What we have is a free string theory with no 
interactions. To model the real world we have to put in interactions. 
Usually the way in which interactions are added in a free field theory is 
by adding nonlinearities in the action. 
The way string theory is formulated, things are not so 
simple because the Polyakov action is not a spacetime action, but a 
worldsheet action. And it is this worldsheet action that tells us things 
about spacetime. 

One consistent way in which we know how to add interactions in string 
theory (perturbatively) is by declaring that the spacetime S-matrix in 
string theory should be defined by the sum of correlation functions in the 
2-dimensional quantum field theory (for example defined by the Polyakov 
action) on the worldsheet - the sum being over all possible topologies 
and all possible distinct metrics\footnote{What do we mean by distinct 
metrics? We obviously do not choose to distinguish between 
metrics that are related by coordinate transformations. In fact we also do 
not choose to distinguish between metrics related by Weyl transformations. 
So the ``distinct" metrics on the surface are metrics which cannot be 
made the same by a coordinate transformation plus Weyl rescaling.} on the 
worldsheet. It turns out that this is a consistent way to add 
interactions, but the problem is that this is a perturbative definition. 

What is so great about a non-perturbative definition of string theory? 
Well, first of all, note that when I quantized the string, I worked in 
flat spacetime. In fact I could have worked in a more general background, 
but I would have preferred my theory to tell me what is the background 
than having to put it in by hand. Usually in field theory we have a 
spacetime action functional formulation of the theory and we can use it to 
calculate the quantum effective action etc., and they can be used to 
calculate what are the acceptable backgrounds (vacua) of the theory. But 
as I said, in string theory we do not have a spacetime formulation (only a 
worldsheet one) and so we do not know how to get this kind of information. 
We really do need new ideas to go further with string theory. This is the 
crux of the statement that we do not have a non-perturbative background 
independent definition of string theory. We do not know how the vacua are 
selected.

I said that we can work in other backgrounds, not just in flat spacetime. 
It turns out that the quantum consistency of string theory imposes 
conditions on our choice of background. One of these conditions is that 
the background metric has to satisfy Einstein's field equation. This is 
the way 
that Einstein's gravity emerges in string theory. 

The bosonic string theory that we formulated in the earlier sections has 
certain obvious problems. First and foremost, we need fermions 
because we know matter is made of fermions. We also saw that the 
ground-state is a tachyon, tachyons signify instabilities in field theory.
The solution to both these problems is to make the theory supersymmetric. 
Supersymmetry is a generalization of Lorentz symmetry that connects 
half-integral and integral spin particles. Supersymmetry manages to remove 
the tachyon and at the same time brings in fermions into the theory. The 
price to pay is that in the real world we do not observe supersymmetry. So 
we need to think of a good way to break supersymmetry in string theory. 
This is again a non-trivial problem. 

Another problem is that we know of many ways to construct superstring 
theories. How do we know which of these is the more fundamental 
description? Well, 
this was a major stumbling block in the development of string theory until 
1995. Then, Edward Witten pointed out that all these string theories are 
related by what are called dualities, and also that all these theories are 
limits of a certain (as yet) unknown theory, tentatively called M-theory 
(for Magic, Mystery or Matrix according to taste.). M-theory is believed 
to be the non-perturbative theory which is supposed to solve all the 
heart-aches of string theorists. But as of now, we have only had a few 
glimpses at M-theory and a full understanding is lacking. 

In superstring theories, the critical 
dimension is 10 as already mentioned. Since we live in a world with 4 
large dimensions, it is thought that the extra six dimensions are wrapped 
up into extremely small sizes. Certain broad restrictions from the 
physics of the standard model can be used to argue that the shapes into 
which 
these extra-dimensions are ``compactified" are what are called Calabi-Yau 
manifolds. The geometry of Calabi-Yau's has a lot to do with the 
properties of the particles that we see in the standard model. CYs are of 
tremendous interest to pure mathematicians and string theory seems to give 
a new handle on them. Actually, some string-inspired ideas were 
instrumental in the discovery of a certain relationship between apparently 
unrelated CY manifolds. This connection, called Mirror symmetry, was 
a great source of excitement to the mathematicians.

This brings us to another issue: that string theory and related 
ideas seem to bring up new ideas in mathematics. This is very unlike the 
usual trend in physics, were usually physicists borrow the 
tools that they need from the already constructed arsenal of 
mathematicians. But string theory seems to necessitate really new ideas in      
mathematics as well as physics. This is one of the things that makes 
string theory hard and exciting at the same time. 

So it seems to me that we are indeed living in a time when there is no 
lack of 
problems in string theory and there is a lot of room for new ideas. I 
thank the organizers once again for letting me introduce this remarkable 
subject to a warm 
audience.  

\section{Acknowledgments}

I would like to thank Prof. Willy Fischler for being a source 
of inspiration. 
Hyuk-Jae Park read parts of the manuscript and offered critical 
comments and I am indebted to him for that.

\end{document}